\documentclass[final,sort&compress]{elsarticle}
\usepackage{hyperref,amsmath} 

\journal{arXiv}

\begin{document}

\begin{frontmatter}

\title{Primary and secondary components of nerve signals}



\author{J\"uri Engelbrecht\corref{mycorrespondingauthor}}
\cortext[mycorrespondingauthor]{Corresponding author}
\ead{je@ioc.ee}

\author{Kert Tamm\corref{cor2}} 
\ead{kert@ioc.ee}
\author{Tanel Peets\corref{cor2}} 
\ead{tanelp@ioc.ee}

\address{Laboratory of Solid Mechanics, Department of Cybernetics, School of Science,\\
Tallinn University of Technology,\\ Akadeemia tee 21, Tallinn 12618, Estonia}

\begin{abstract}
The action potential propagating in a nerve fibre generates accompanying mechanical and thermal effects. The whole signal is therefore an ensemble which includes primary and secondary components. The primary components of a signal are the action potential itself and longitudinal mechanical waves in axoplasm and surrounding biomembrane. These components are characterized by corresponding velocities. The secondary components of   a signal are derived from primary components and include transverse displacement of a biomembrane and the temperature -- these have no independent velocities but have been measured in several experiments. A robust mathematical model is presented based on differential equations describing the signal primary components which are coupled into a system by coupling forces. The model includes also mathematical formulation for establishing the secondary components following the ideas from experimental studies. 
\end{abstract}


\end{frontmatter}


\section{Introduction}
The main structural element of a nervous system is the axon which ensures the signal propagation from a nerve cell to a nerve terminal. Axons are built as cylindrical biomembranes made of lipid bilayers and filled with a viscous fluid called axoplasm (intracellular fluid). This structure called also a nerve fibre, is placed into a surrounding extracellular (intersticial) fluid. The electrical signal in the axoplasm is called action potential (AP). The contemporary understanding of axon physiology is described in many overviews \cite{Nelson2004,Clay2005,Debanne2011} etc. 
Beside the electrical signals which are supported by ionic currents through the biomembrane, there are also accompanying effects: deformation of the biomembrane, pressure in the axoplasm, temperature changes, etc.
In this paper, an attempt is presented to build up a mathematical model able to describe the propagation of an AP together with main accompanying effects.
\section{The physiological background}
The AP and the accompanying effects form actually an ensemble of waves. The general signal propagation can be described in a following way. The action potential (AP) is generated by an electrical impulse with an amplitude above the threshold. The AP propagates in the axoplasm supported by ion currents through the biomembrane. The AP also generates a pressure wave (PW) in the axoplasm and a longitudinal wave (LW) in the biomembrane. The LW means a local longitudinal compression which changes the diameter of the biomembrane (swelling) reflected by a transverse wave (TW). As a result of this complicated process, the temperature change accompanying the signal propagation is also measured. In the ensemble of waves, the profile of the AP depends on the properties of the axoplasm and ion currents, the PW depends on the properties of the axoplasm. Both the LW and TW depend on the elastic properties of the biomembrane. 
These waves are characterized by the following values of their physical parameters. First, the AP has a resting potential in the axoplasm about $-60$ to $-80$ mV, AP increase (amplitude) is about $100-110$ mV, its duration is typically about  $1$ ms or in some cases even more (without the refractory time), its velocity depends on the diameter and varies from $2$ to $110$ m/s (estimates by many authors), the AP has an overshoot responsible for the refractory period. Second, the PW has a value of about 1 to 10 mPa \cite{Terakawa1985}. In addition, the TW has an amplitude about $1-2$ nm \cite{Terakawa1985,Tasaki1988}. Based on measurements, the AP has an asymmetric shape with an overshoot with respect of the resting potential, the PW has a unipolar shape with a possible overshoot, the TW has a bipolar shape which corresponds to a unipolar LW. The temperature change during the passage of the AP is measured about 20 to 30 $\mathrm{\mu}$K$^{\circ}$ \cite{Tasaki1988} for garfish but also much less for bullfrog \cite{Tasaki1992}. However, the experimental results under various conditions may differ from these estimations.  

\section{Mathematical modelling of nerve signals}
There has been a considerable interest to build up a mathematical model which could grasp all (or most essential) effects of signals in nerve fibres. The models for single waves are known. The celebrated Hodgkin-Huxley model \cite{Hodgkin1945} (shortly HH model) which describes the propagation of the AP, has been fundamental for many studies. The simplified FitzHugh-Nagumo (FHN) model considers just one ionic current compared with three in the HH model but describes also an asymmetric pulse with an overshoot \cite{Nagumo1962}. The PW can be modelled according to well-known theory of fluids – either by Navier-Stokes equations or taking into account its very small intensity - by the classical wave equation with viscosity included. A mathematical model for the LW in biomembranes is proposed by Heimburg and Jackson \cite{Heimburg2005} and improved by Engelbrecht et al \cite{EngelbrechtTammPeets2014}. 

The basic question is to understand the electromechanical transduction of energy from the AP to other waves. It is argued that either electrostriction or piezoelectricity is the main mechanism of transduction \cite{Gross1983}. In the process of electrostriction, the produced stress is proportional to the square of the imposed electric field but the question is still open. It seems that the field value of an electrical field determines the static situation. It has also been stated \cite{Gonzalez-Perez2016} that the mechanical changes in biomembrane are proportional to voltage changes.
As far as the LW and TW are mechanical effects then the knowledge from classical mechanics turned to be useful: following the theory of rods \cite{Porubov2003}, the TW is related to the space derivative of the LW, i.e., to the compressibility of the biomembrane. 
In addition to mechanical effects, the heat production and temperature changes accompany the signal propagation in nerve fibers \cite{Abbott1958,Howarth1968,Heimburg2008,Ritchie1985}. There are several proposals for explaining this physical process (see below). 

Once the mathematical models of single waves or processes are known, it seems to be straightforward to unite them into a coupled system. Here, however, the crucial question arises: what are the coupling forces? In other words: what are the coupling mechanisms? The mathematical modelling has a hidden strength -- it is not only a tool for simulations but also a tool to establish the structure of a process under investigation and to understand causality. 

\section{A coupled model of the nerve signalling }
\begin{figure}[h]
\includegraphics[width=0.45\textwidth]{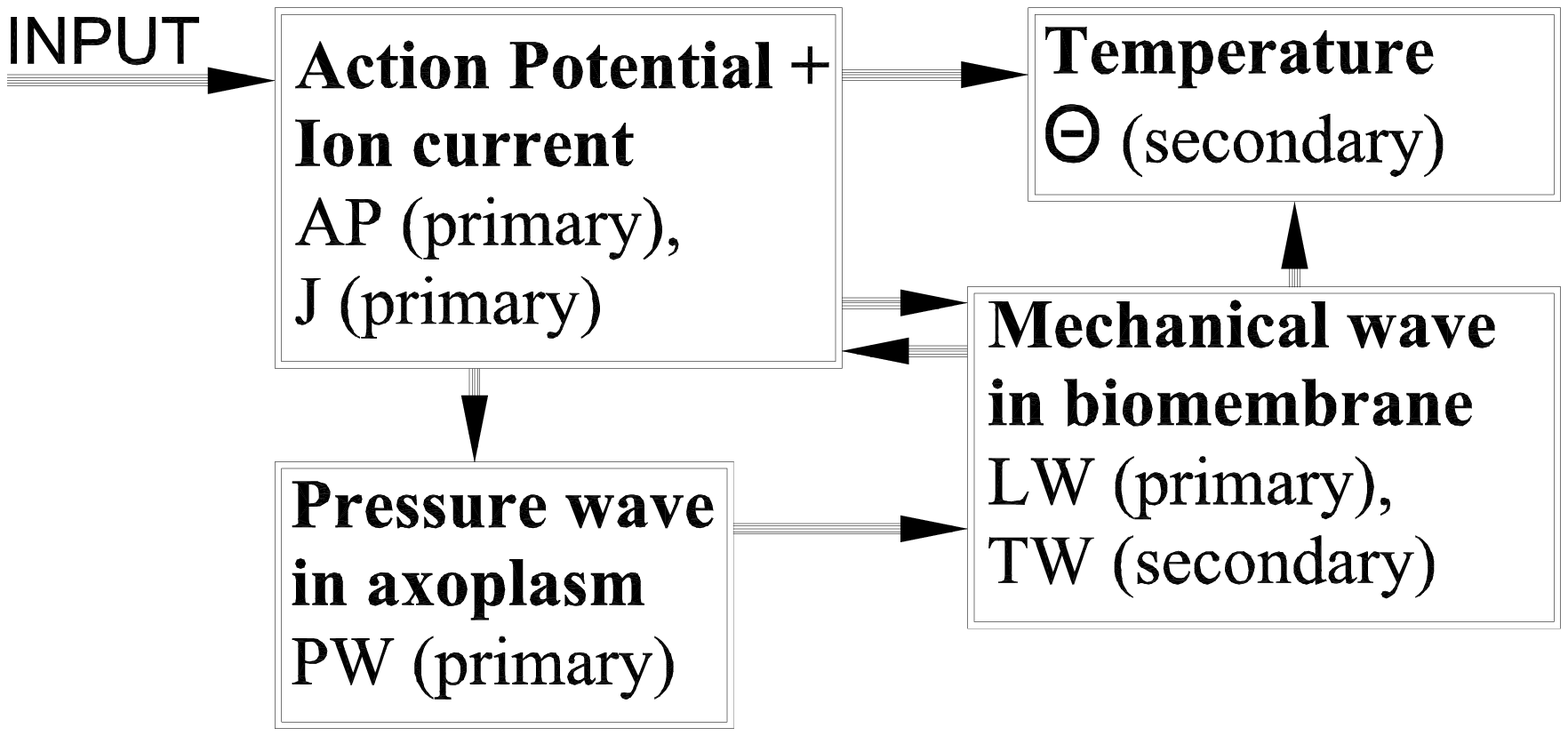}
\includegraphics[width=0.54\textwidth]{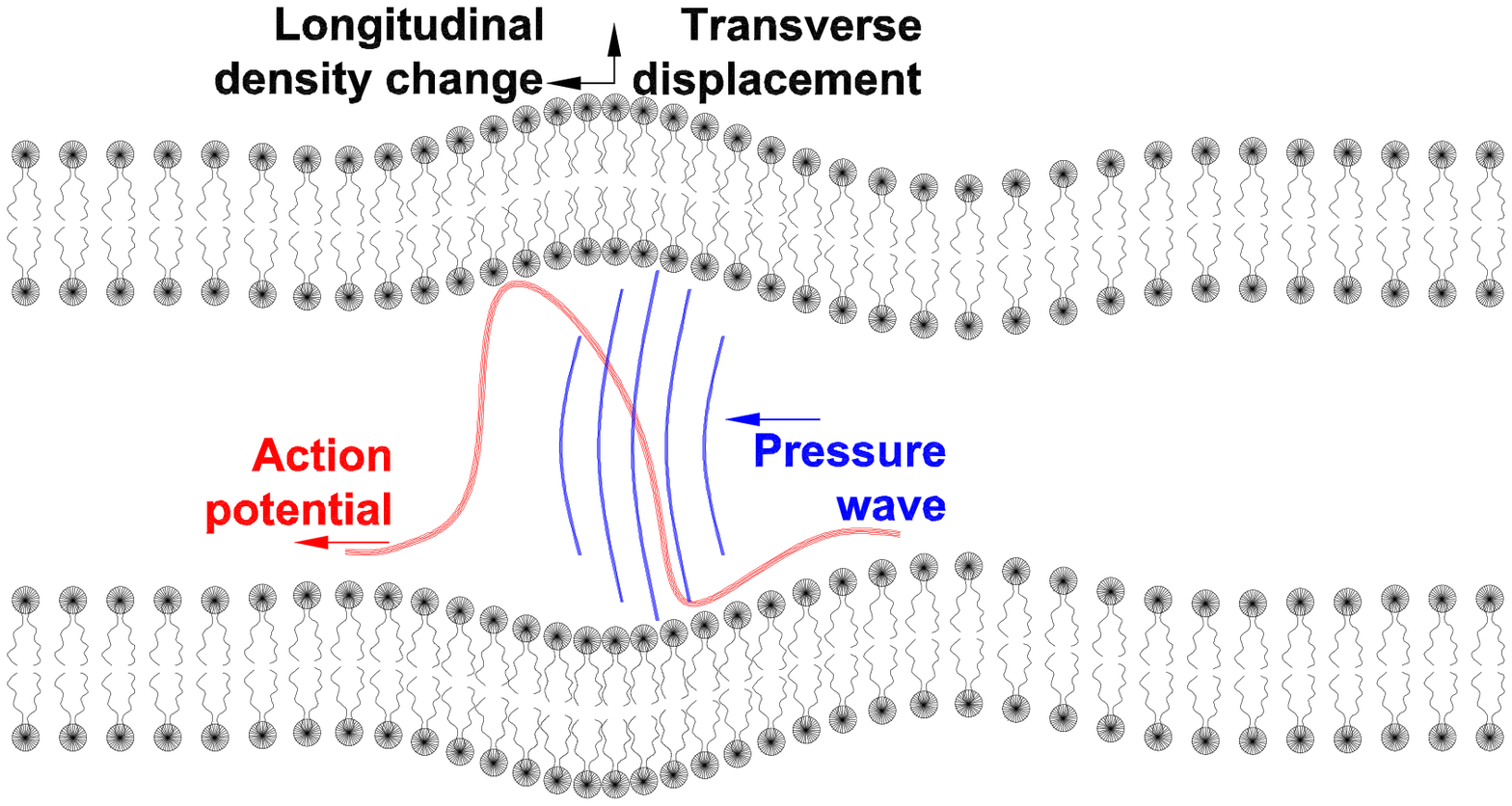}
\caption{The block diagram of the possible mathematical model for the ensemble of waves in the nerve fiber (left) and an artistic sketch (not in scale) of wave ensemble propagation in the axon (right).}
\label{Fig1}
\end{figure}

In general terms, the wave ensemble in a nerve fibre is composed by \emph{primary} components: the AP, the PW, and the LW. These waves when considered separately possess finite velocities but in the ensemble, the velocities are synchronized. The synchronization demonstrated in experiments is a process which needs further analysis. The \emph{secondary} (i.e., derived) components are the TW and temperature change $\Theta$.  These components have no independent velocities and cannot emerge without primary components. In this way, the main physical properties of a fibre are taken into account by the primary components. The block-scheme of signal generation with primary and secondary components of an ensemble of waves is shown in Fig.~\ref{Fig1}.

The primary components of the ensemble are coupled through coupling forces. Our hypothesis is the following: the mechanical waves in axoplasm and the surrounding biomembrane are generated due to changes in electrical signals \cite{Engelbrecht2018c,Engelbrecht2018d}. In mathematical terms, the changes are either the space or time derivatives. As far as the electrical signals (either the AP or ionic currents) have a pulse-like shape then their derivatives are of the bipolar shape which is energetically balanced. In the first approximation, the forces can be introduced by the first-order polynomials. The proof of concept with using non-dimensionalized equations has shown \cite{Engelbrecht2018c,Engelbrecht2018} that the calculated wave profiles correspond qualitatively to measured ones as seen from Fig.~\ref{Fig2}. The equations used and the parameter values for calculations are described in the Appendix. In this way, the scene is set but the crucial problem is to quantify the coupling forces.

\begin{figure}[h]
\includegraphics[width=0.49\textwidth]{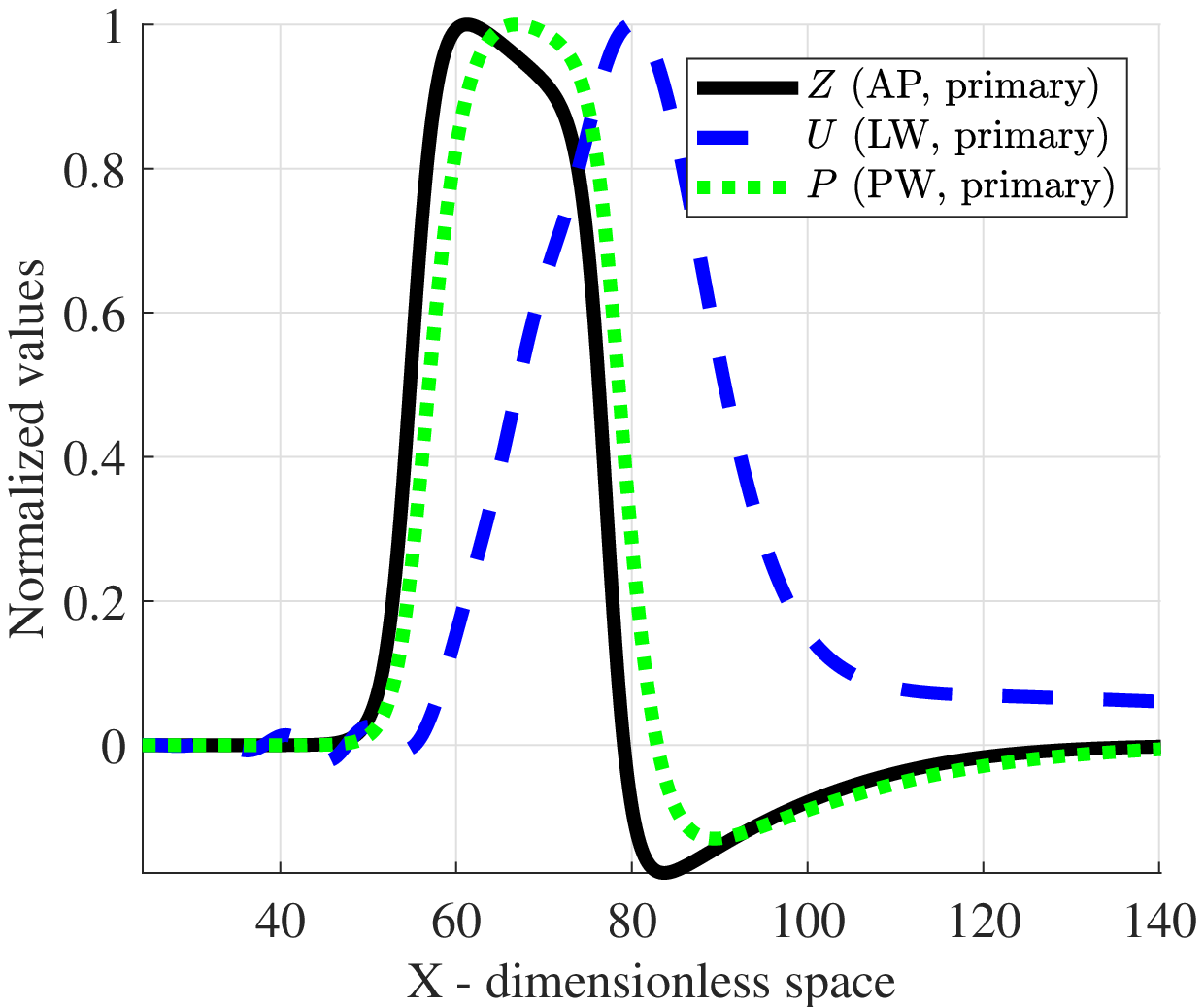}
\includegraphics[width=0.49\textwidth]{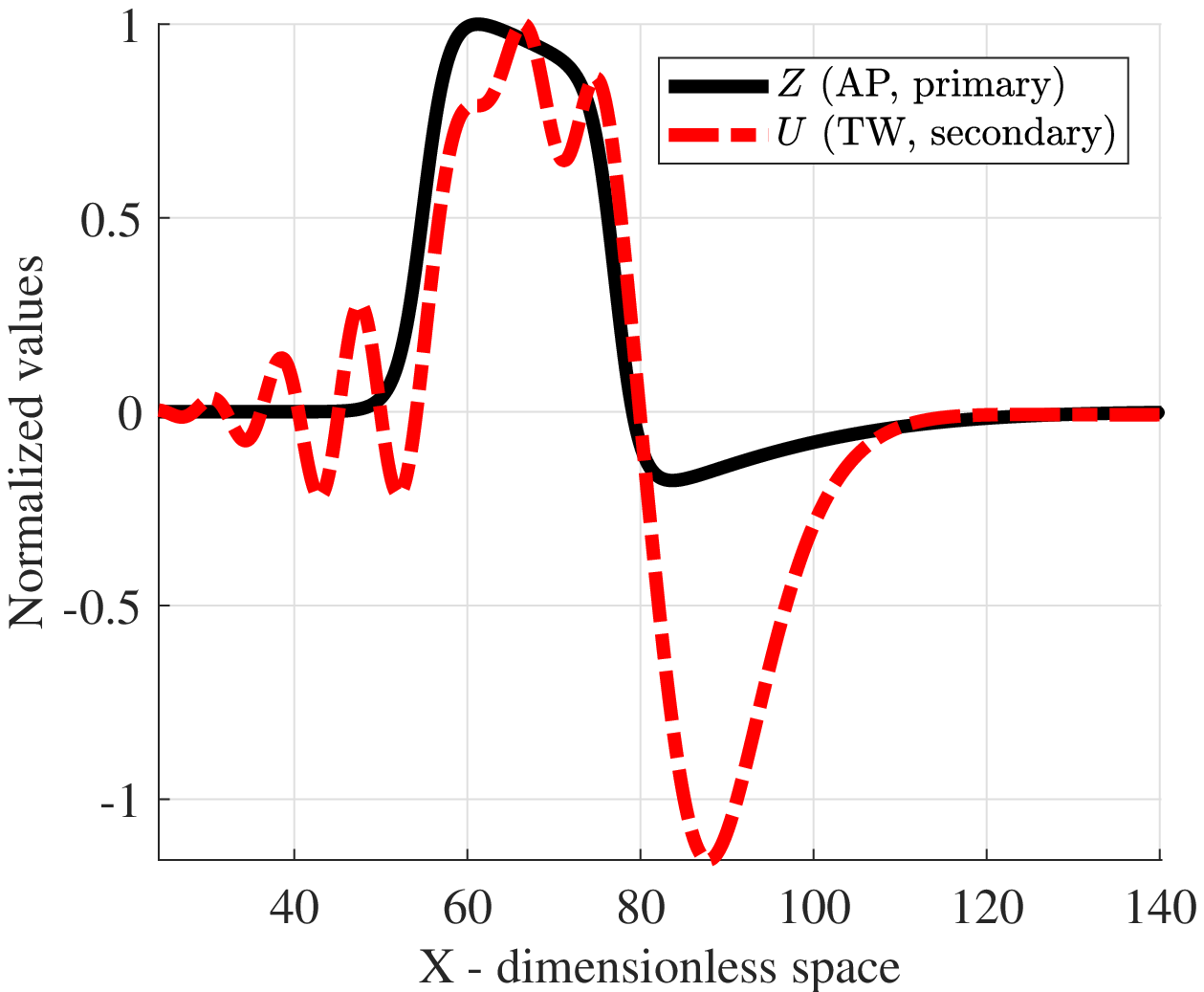}
\caption{The ensemble of the primary components AP, the PW and LW (left panel), the secondary component TW (right panel). Parameter values in Appendix.}
\label{Fig2}
\end{figure}

The secondary components of a nerve signal also need proper modelling. As said before, the TW can be calculated knowing the profile of the LW in the biomembrane by the space derivative following the ideas of solid mechanics of rods. Following this idea \cite{Engelbrecht2018}, the result of mathematical simulation is qualitatively similar to experimental results in phase of the AP \cite{Terakawa1985,Yang2018} -- see Fig.~\ref{Fig2} and Appendix.


The mechanism of temperature changes is more complicated and needs more explanation. There are many measurements demonstrating the temperature changes starting from earlier studies \cite{Downing1926,Abbott1958,Howarth1968} to results of \cite{Tasaki1988} and  \cite{Tasaki1992}, see also \cite{Andersen2009}. However, the clear understanding about the mechanism of the heat production is still under discussion.

\begin{figure}[h]
\includegraphics[width=0.49\textwidth]{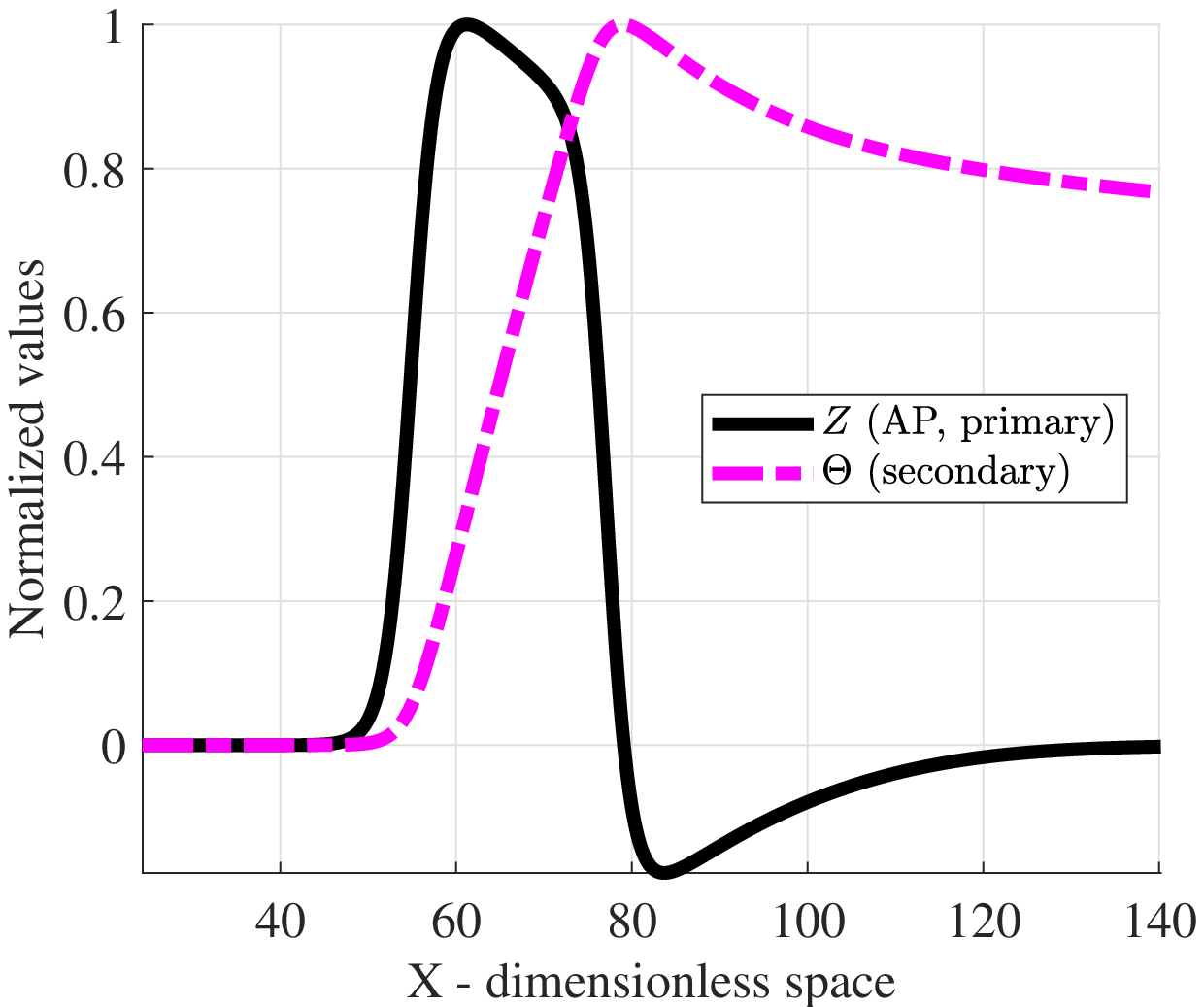}
\includegraphics[width=0.49\textwidth]{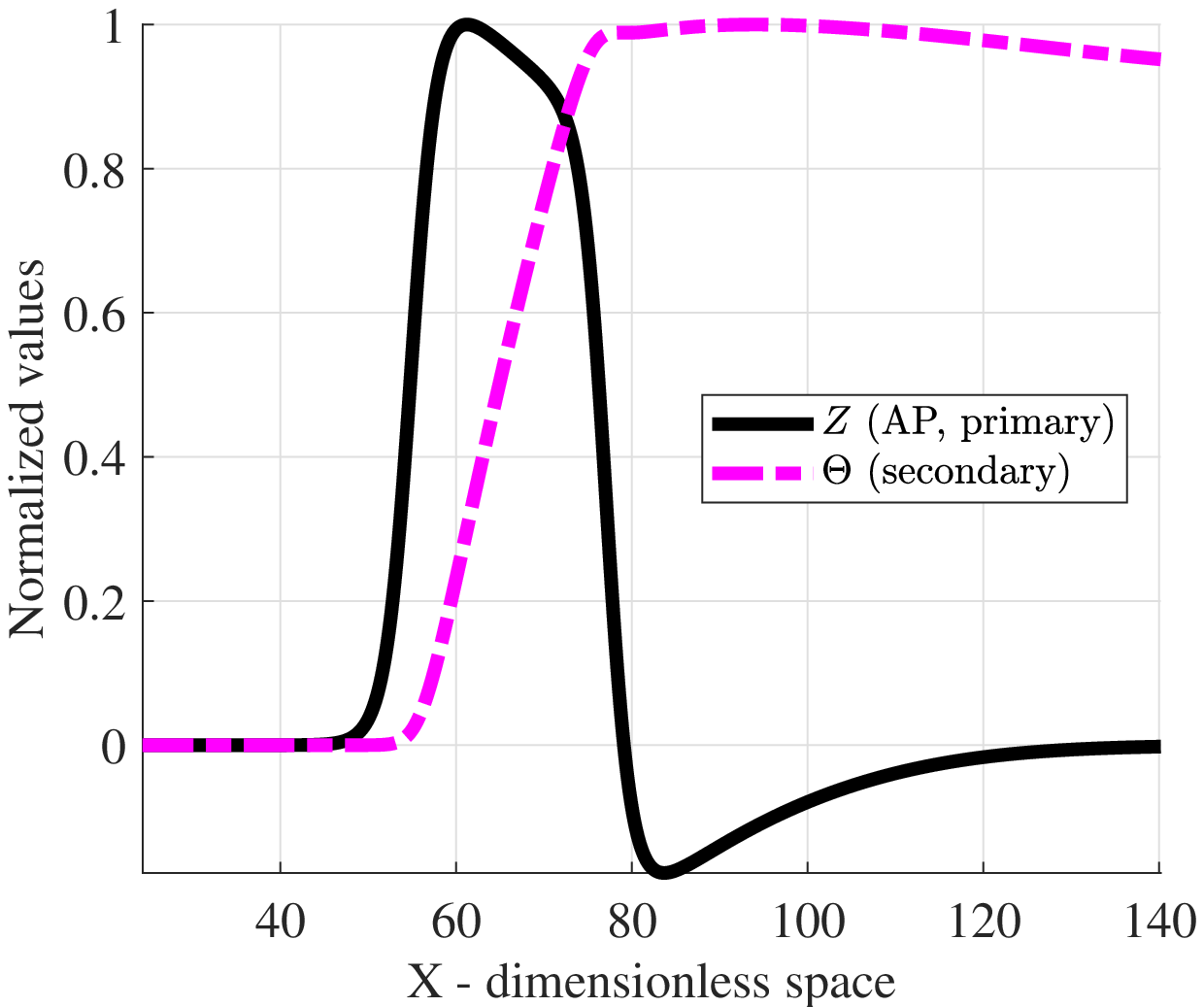}
\caption{The temperature increase inside the axon in the case of heat generation proportional to $Z$ (left panel) and proportional to $Z^{2}$ (right panel). Parameter values in Appendix.}
\label{Fig4}
\end{figure}

\begin{figure}[h]
\includegraphics[width=0.49\textwidth]{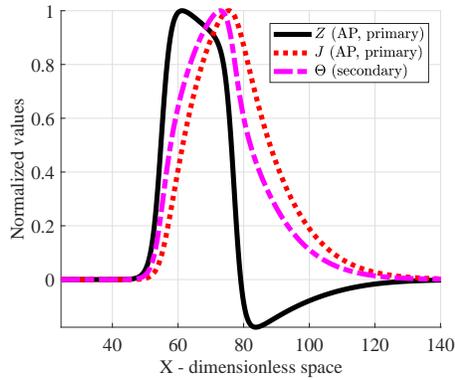}
\caption{The temperature increase inside the axon in the case of heat generation proportional to $\tau_1 Z_T + \tau_2 J_T$. Parameter values in Appendix.}
\label{Fig5}
\end{figure}

However, several proposals for simulation the temperature changes could be cast into mathematical models. First, energy released depends on the voltage square \cite{Heimburg2008,Ritchie1985}  and then temperature depends on the integral of voltage square. Second, depending on time constants (RC:  resistance x capacitance), temperature may be roughly proportional to voltage or temperature rate may be proportional to voltage \cite{Tasaki1992}. 
The general model for temperature changes can be derived by using the Fourier law and energy conservation together with assumptions on the possible sources. Then temperature $\Theta$ is governed by:
\begin{equation}
\label{heateq}
\frac{\partial\Theta}{\partial T} = \alpha  \frac{\partial^2 \Theta}{\partial X^2} + F_3(Z,J,U),
\end{equation}
where $Z$ is the AP, $J$ is the ion current, $U$ is the density change, $\alpha$ is coefficient and $F_3$ is the source term. 
Some corresponding simulation results based on this model are shown in Figs \ref{Fig4} and \ref{Fig5} following \cite{Tamm2018}. Equations and numerical values of parameters used in simulations are presented in the Appendix.

\section{Final remarks}
The coupled model described above is certainly a robust one. Dealing with the primary components, the main condition for the proof of concept is to have an AP in a proper form that is why the simple FHN model is used within this analysis. There are several possible modifications (and complications) of this model to bring the description of signalling closer to realities. Certainly the HH model could give more detailed description of an AP with many additional ion current involved. The existence of several ion currents may explain the retardation of refraction and the influence of anesthetics. The viscosity and existing filaments in axoplasm could be of importance for the generation of the PW and ion currents. The inhomogeneities of the biomembrane may play significant role for opening the ion channels and in the process of forming the stationary AP. So the chemical and molecular effects accompanying signalling must be analysed in order to improve modelling \cite{Tasaki1988,HJ2007}. 

For modelling the secondary components, the mechanism for the heat production needs well-grounded clarification.
Nevertheless, the proposal that source term $F_3$ in Eq.~\eqref{heateq} depends on $Z$ or $Z^2$ (see Fig.~\ref{Fig4}) permits qualitatively  to match experimental results by Abbott et al \cite{Abbott1958} while using the derivatives of the amplitude of the AP and the ion current, the temperature change is qualitatively similar to experiments by Tasaki and Byrne \cite{Tasaki1992}. It cannot be excluded that there exist several mechanisms for generating the temperature changes accompanying the primary components of signal in a  nerve fibre.

Despite of numerous experiments and general knowledge about axon physiology, there are still many challenges in building general models of nerve signal propagation. The separation of the whole process into primary and secondary components permits to distinguish wave-like processes from accompanying secondary effects which certainly play also an important role in the whole process. In this context the energy balance must be studied in more detail.

\section*{Acknowledgments}
This research was supported by the European Union through the European Regional Development Fund (Estonian Programme TK 124) and by the Estonian Research Council (projects IUT 33-24, PUT 434).

\appendix
\section{}
The used model equations, the numerical scheme and the parameter values used in the calculations are shortly summarized below. 
Here and further index $X$ denotes spatial partial derivative and index  $T$ denotes partial derivative with respect to time. The model equations are the following -- the primary components:
\begin{equation}
\label{FHNeq}
\begin{split}
& Z_{T} = D Z_{XX} + Z \left( Z - \left[ a_1 + b_1 \right] - Z^2 + \left[ a_1 + b_1 \right] Z \right) - J, \\
& J_{T} = \varepsilon \left( \left[ a_2 + b_2 \right] Z - J \right),
\end{split}
\end{equation}
\begin{equation}
\label{iHJeq}
\begin{split}
U_{TT} = \, & c^2 U_{XX} + N U U_{XX} + M U^2 U_{XX} + N U_{X}^{2} + \\
& + 2 M U U_{X}^{2} - H_1 U_{XXXX} + H_2 U_{XXTT} + F_1(Z,J,P),
\end{split}
\end{equation}
\begin{equation}
\label{Peq}
P_{TT} = c_{f}^{2} P_{XX}  - \mu P_T + F_2(Z,J,U);
\end{equation}
the secondary components:
\begin{equation}
\label{TWeq}
W = k U_{X},
\end{equation}
\begin{equation}
\label{Heq}
\Theta_{T} = \alpha \Theta_{XX} + F_3(Z,J,U),
\end{equation}
where $Z$ is potential, $J$ is the abstracted ion current (characteristic to the FHN model), $a_i, b_i$ are ``electrical" and ``mechanical" activation coefficients, $D, \varepsilon$ are coefficients, $U$ is longitudinal density change, $c$ is sound velocity of unperturbed state in the lipid bi-layer, $N, M$ are nonlinear coefficients, $H_1, H_2$ are dispersion coefficients, $P$ is pressure, $c_{f}$ is sound velocity in axoplasm, $\mu$ is dampening coefficient, $W$ is the transverse displacement of the biomembrane, $k$ is coefficient (in theory of rods related to Poisson ratio \cite{Porubov2003}), $\Theta$ is the temperature  and $\alpha$ is the combined coefficient determining the diffusion speed for thermal energy in the environment. 

Coupling forces 
\begin{equation}
\label{Forces}
F_1 = \gamma_1  P_T + \gamma_2 J_T - \gamma_3 Z_T;
\quad
F_2 = \eta_1 Z_X + \eta_2 J_T + \eta_3 Z_T,
\end{equation}
where $\gamma_i$ are coupling coefficients for mechanical wave and  $\eta_i$ are coupling coefficients for the pressure wave. Time derivatives act across membrane and spatial derivatives act along the axon.

In the present paper three different functions $F_3$ have been used for generating and consuming thermal energy in the \eqref{Heq} under different example scenarios which are
\begin{equation}
F_3 = \tau_1 Z; \quad F_3= \tau_2 Z^{2}; \quad F_3= \tau_3 Z_T + \tau_3 J_T,
\end{equation}
where $\tau_i$ are the thermal coupling coefficients. 
The model equations are solved using the pseudospectral method (PSM), see \cite{Engelbrecht2018d} and references therein. 

The parameters have the following values in the present paper: 
\begin{enumerate}[(i)]
\item for the FHN equation \eqref{FHNeq} \\
$
D=1;
\varepsilon=0.018;
a_1=0.2;
a_2=0.2;
b_1=-0.05 \cdot U;
b_2=-0.05 \cdot U;
$
\item for the improved Heimburg-Jackson model \eqref{iHJeq} \\
$
c^2 = 0.10;
N=-0.05;
M=0.02;
H_1=0.2;
H_2=0.99;
$
\item for the  pressure \eqref{Peq} \\
$
c_{f}^{2}=0.09;
\mu=0.05;
$
\item for the transverse displacement \eqref{TWeq} and heat equation \eqref{Heq} \\
$
k=1; \alpha=0.05;
$ 
\item and the coupling coefficients are \\
$
\gamma_1=0.008; 
\gamma_2=0.01; 
\gamma_3=3 \cdot 10^{-5}; 
\eta_1=0.005; 
\eta_2=0.01; 
\eta_3=0.003;\\ 
\tau_1 = 5 \cdot 10^{-5};
\tau_2 = 5 \cdot 10^{-5}; 
\tau_{3} = 5 \cdot 10^{-5};
\tau_4=1 \cdot 10^{-3}.
$
\end{enumerate}
In addition the parameters for the numerical scheme are $L=64\pi$ (the length of the spatial period), $T_f = 400$ (the end time for integration, figures in the present paper are shown at $T=400$), $n=2048$ (the number of spatial grid points), $A_z=1.2$ (the initial amplitude for the $Z$), $B_o=1.0$ (the width of the initial sech$^2$ type pulse for $Z$).


\end{document}